\documentclass[twocolumn,nofootinbib]{revtex4-1}
\usepackage{amsmath,amssymb,bm,graphicx,hyperref}

\newcommand\grad{\bm\nabla}
\newcommand\+{\dagger}
\newcommand\<{\langle}
\renewcommand\>{\rangle}
\newcommand\up{\uparrow}
\newcommand\down{\downarrow}
\newcommand\eps{\epsilon}
\newcommand\0{{\bm{0}}}
\newcommand\A{{\bm{A}}}
\newcommand\Q{{\bm{Q}}}
\newcommand\x{{\bm{x}}}
\newcommand\z{{\bm{z}}}
\newcommand\N{\mathbb{N}}
\newcommand\FF{\mathrm{FF}}
\newcommand\LO{\mathrm{LO}}
\newcommand\MF{\mathrm{MF}}

\begin{document}

\title{Two-dimensional imbalanced Fermi gas in antiparallel magnetic fields}

\author{Takaaki Anzai}
\author{Yusuke Nishida}
\affiliation{Department of Physics, Tokyo Institute of Technology,
Ookayama, Meguro, Tokyo 152-8551, Japan}

\date{July 2019}

\begin{abstract}
We study a two-dimensional Fermi gas with an attractive interaction subjected to synthetic magnetic fields, which are assumed to be mutually antiparallel for two different spin components with population imbalance.
By employing the mean-field approximation, we show that the Fulde-Ferrell state is energetically favored over the Larkin-Ovchinnikov state in the weak-coupling limit.
We then elucidate the zero-temperature phase diagram in the space of attraction and two chemical potentials analytically at weak coupling as well as numerically beyond it.
Rich structures consisting of quantum Hall insulator, unpolarized superfluid, and Fulde-Ferrell phases separated by various second-order and first-order quantum phase transitions are found.
\end{abstract}

\maketitle

\section{Introduction}
Ceaseless progress in ultracold atom experiments allows us to control system parameters at will, such as interaction, dimensionality, statistics, and internal degrees of freedom of atoms~\cite{Bloch:2008}.
Furthermore, it became possible to apply synthetic magnetic fields to neutral atoms by optically coupling their internal states, so that quantum phenomena induced by magnetic fields are also within the reach of experimental realization~\cite{Dalibard:2011,Goldman:2014,Cooper:2019}.
This approach was further extended to create ``antiparallel'' magnetic fields, which act on two different spin components of atoms with the same magnitude but in opposite directions~\cite{Beeler:2013,Aidelsburger:2013,Kennedy:2013}.

Motivated by such experimental abilities to control the interaction, dimensionality, and magnetic fields, the authors of this paper previously studied a two-dimensional (2D) Fermi gas with an attractive interaction between two spin components in antiparallel magnetic fields~\cite{Anzai:2017}.
Here, its phase diagram at zero temperature was found to show the rich structure consisting of pair superfluid and quantum spin Hall insulator phases, which are separated by a second-order quantum phase transition classified into the universality class of either the dilute Bose gas or the XY model.
For related theoretical works, see Refs.~\cite{Liu:2009,Fialko:2014,Furukawa:2014,Yoshino:2019,Feng:2015} on 2D Bose and three-dimensional (3D) Fermi gases as well as Refs.~\cite{Cocks:2012,Wang:2014,Peotta:2015,Iskin:2015,Umucalilar:2017,Iskin:2017,Iskin:2018,Zeng:2019} in optical lattices.
The purpose of this paper is to extend our previous analysis on the zero-temperature phase diagram to the population imbalanced system with particular attention on the possible Fulde-Ferrell (FF) and Larkin-Ovchinnikov (LO) phases~\cite{Conduit:2008,Yin:2014,Sheehy:2015,Toniolo:2017}.

The FF state is an anisotropic superfluid state where the order parameter has a spatially varying phase with a constant magnitude, $\Delta(\x)=e^{i\Q\cdot\x}\Delta$, so that the Cooper pairing takes place with nonzero momentum~\cite{Fulde:1964}.
On the other hand, the LO state is an inhomogeneous superfluid state where the order parameter is periodically modulated in its magnitude, such as $\Delta(\x)=\cos(\Q\cdot\x)\Delta$~\cite{Larkin:1965}.
Although the FF state is often assumed by an ansatz because of its ease to handle theoretically, the LO state has been known to be energetically favored in the familiar population imbalanced systems~\cite{Matsuo:1998,Shimahara:1998,Yoshida:2007,Bulgac:2008,Toniolo:2017}.
In contrast, when the antiparallel magnetic fields are applied, we will find below that the FF state turns energetically favored at least in the weak-coupling limit where the mean-field (MF) approximation may be employed.
After formulating the mean-field Hamiltonian in Sec.~\ref{sec:mean-field}, we will elucidate the zero-temperature phase diagram in the space of attraction and two chemical potentials analytically at weak coupling in Sec.~\ref{sec:weak-coupling} as well as numerically beyond it in Sec.~\ref{sec:beyond}.

\section{Mean-field Hamiltonian}\label{sec:mean-field}
The system under consideration is composed of spin-$1/2$ fermions in 2D subjected to spin-dependent vector potentials, whose Hamiltonian reads
\begin{align}
H &= \sum_{\sigma=\up,\down}\int\!d\x\,\phi_\sigma^\+(\x)
\left[-\frac{[\grad+i\A_\sigma(\x)]^2}{2m}-\mu_\sigma\right]\phi_\sigma(\x) \notag\\
&\quad - g\int\!d\x\,\phi_\up^\+(\x)\phi_\down^\+(\x)\phi_\down(\x)\phi_\up(\x).
\end{align}
Here, we set $\hbar=1$, $m$ is the mass of fermions, $\mu_\sigma$ is the chemical potential for each spin component, and the coupling constant $g>0$ is assumed to be attractive.%
\footnote{We note that $mg/\hbar^2$ is dimensionless in 2D and is related to the 3D scattering length via $mg/\hbar^2=-\sqrt{8\pi}\,a_\mathrm{3D}/\ell_z$ within the Born approximation, where $\ell_z\equiv\sqrt{\hbar/m\omega_z}\gg-a_\mathrm{3D}>0$ is a transverse harmonic-oscillator length~\cite{Bloch:2008}.}
We also choose the vector potentials as
\begin{align}
\A_\up(\x) = -\A_\down(\x) = -By\hat\x,
\end{align}
so that different spin components experience antiparallel magnetic fields with the magnitude $B>0$; $\grad\times\A_\up(\x)=-\grad\times\A_\down(\x)=B\hat\z$.
The standard mean-field approximation leads to
\begin{align}\label{eq:mean-field}
& H_\MF = \sum_{\sigma=\up,\down}\int\!d\x\,\phi_\sigma^\+(\x)
\left[-\frac{[\grad+i\A_\sigma(\x)]^2}{2m} - \mu_\sigma\right]\phi_\sigma(\x) \notag\\
& + \int\!d\x\left[\frac{|\Delta(\x)|^2}{g} - \Delta^*(\x)\phi_\down(\x)\phi_\up(\x)
- \phi_\up^\+(\x)\phi_\down^\+(\x)\Delta(\x)\right],
\end{align}
where $\Delta(\x)=g\<\phi_\down(\x)\phi_\up(\x)\>$ is the superfluid order parameter.

We recall that the eigenfunction of the single-particle Hamiltonian in the Landau gauge for $\sigma=\,\up$ is
\begin{align}\label{eq:eigenfunction}
\chi_{kl}(x) \equiv \frac{e^{ikx}}{\sqrt{L}}\,F_l(y-k\ell_B^2)
\end{align}
with
\begin{align}
F_l(y) \equiv \frac{e^{-(y/\ell_B)^2/2}}{\sqrt{2^ll!\,\pi^{1/2}\ell_B}}\,
H_l\!\left(\frac{y}{\ell_B}\right)
\end{align}
being the $l$th eigenfunction of the harmonic oscillator~\cite{Landau-Lifshitz}.
It solves the Schr\"odinger equation $[-(\grad-iBy\hat\x)^2/(2m)]\chi_{kl}(x)=\eps_l\chi_{kl}(x)$ with the single-particle energy provided by $\eps_l\equiv(l+1/2)\omega_B$, where $l=0,1,2,\dots$ labels Landau levels.
On the other hand, the eigenfunction for $\sigma=\,\down$ is $\chi_{kl}^*(x)$ with the same single-particle energy.
Here, $\ell_B\equiv1/\sqrt{B}$ and $\omega_B\equiv B/m$ are the magnetic length and the cyclotron frequency, respectively, and $k\equiv2\pi n/L$ is the wave number with $n=0,1,\dots,m\omega_BL^2/2\pi$.
The linear size of the system $L$ is formally kept finite in intermediate calculations, whereas the thermodynamic limit $L\to\infty$ is taken at the end.

\section{Weak-coupling limit}\label{sec:weak-coupling}
\subsection{FF state versus LO state}
In order for the Cooper pairing to take place with an infinitesimal coupling $g\to0$, each chemical potential must lie right at a Landau level, i.e., $\mu_\sigma=\eps_{l_\sigma}$ for some $l_\sigma\in\N_0$ because the system is otherwise insulating~\cite{Anzai:2017}.
When this is the case, the fermion field operator can be expanded over the eigenfunctions in Eq.~(\ref{eq:eigenfunction}) restricted to the $l_\sigma$th Landau level because the mixing with the other Landau levels is negligible in the weak-coupling limit.
By substituting $\phi_\up(\x)=\sum_k\chi_{kl_\up}(\x)\tilde\phi_\up(k)$ and $\phi_\down(\x)=\sum_k\chi_{kl_\down}^*(\x)\tilde\phi_\down(k)$ into Eq.~(\ref{eq:mean-field}), the mean-field Hamiltonian reads
\begin{align}\label{eq:hamiltonian}
H_\MF &= \int\!d\x\,\frac{|\Delta(\x)|^2}{g} \notag\\
& + \sum_{k,k'}\tilde\Phi^\+(k)
\begin{pmatrix}
0 & -\tilde\Delta(k,k') \\
-\tilde\Delta^*(k',k) & 0
\end{pmatrix}
\tilde\Phi(k'),
\end{align}
where the last term is the quasiparticle Hamiltonian in the Nambu-Gor'kov basis with $\tilde\Phi(k)\equiv[\tilde\phi_\up(k),\tilde\phi_\down^\+(k)]^T$ and
\begin{align}
\tilde\Delta(k,k') \equiv \int\!d\x\,\chi_{kl_\up}^*(\x)\Delta(\x)\chi_{k'l_\down}(\x)
\end{align}
measures the overlap of two fermion wave functions with the pair field.
The resulting Hamiltonian allows us to evaluate and compare the ground-state energies for the FF and LO states.

Because of the rotational invariance, the order parameter in the FF ansatz can be chosen to be $\Delta(\x)=e^{iQy}\Delta$, for which we obtain
\begin{align}\label{eq:FF}
\tilde\Delta(k,k') = \delta_{kk'}e^{ikQ\ell_B^2}f_Q\Delta.
\end{align}
Here, the overlap is
\begin{align}
f_Q \equiv e^{-(Q\ell_B)^2/4}\sqrt{\frac{l_\down!}{l_\up!}}
\left(\frac{iQ\ell_B}{\sqrt2}\right)^{l_\up-l_\down}
L_{l_\down}^{l_\up-l_\down}\!\left(\frac{(Q\ell_B)^2}{2}\right),
\end{align}
and we employed
\begin{align}
& \int_{-\infty}^{\infty}\!dx\,e^{-x^2}H_m(x+y)H_n(x+z) \notag\\
&= 2^mn!\,\pi^{1/2}y^{m-n}L_n^{m-n}(-2yz)
\end{align}
with the understanding of $L_n^{m-n}(x)=(m!/n!)(-x)^{n-m}L_m^{n-m}(x)$~\cite{Gradshteyn-Ryzhik}.
Therefore, the eigenvalues of the quasiparticle Hamiltonian in Eq.~(\ref{eq:hamiltonian}) are provided by $\pm|f_Q\Delta|$ for every $k$, so that the ground-state energy is
\begin{align}
\frac{E_\FF}{L^2} = \frac{|\Delta|^2}{g} - \frac{m\omega_B}{2\pi}|f_Q\Delta|.
\end{align}
Its minimization with respect to $\Delta$ and $Q$ leads to
\begin{align}\label{eq:energy_FF}
\min\frac{E_\FF}{L^2} = -g\left(\frac{m\omega_B}{4\pi}|f_{\bar Q}|\right)^2,
\end{align}
where $|\Delta|=g\frac{m\omega_B}{4\pi}|f_{\bar Q}|$ with $\bar Q$ maximizing $|f_Q|$.
Although $\bar Q=0$ for $l_\up=l_\down$ without population imbalance corresponding to the unpolarized superfluid state as expected~\cite{Anzai:2017}, we find $\bar Q\neq0$ for $l_\up\neq l_\down$ with population imbalance corresponding to the FF state.
As examples, $|\bar Q|/\ell_B=\sqrt2$, $2$, and $\sqrt{5-\sqrt{17}}$ are found for $(l_\up,l_\down)=(1,0)$, $(2,0)$, and $(2,1)$, respectively.
We also note that the order parameter shows the remarkable linear dependence on the coupling constant in contrast to the familiar exponential dependence as a consequence of ``magnetic catalysis''~\cite{Miransky:2015,Anzai:2017}.

The same analysis can be repeated in the LO ansatz, $\Delta(\x)=\cos(Qy)\Delta$, for which we obtain
\begin{align}
\tilde\Delta(k,k') = \delta_{kk'}\frac{e^{ikQ\ell_B^2}+(-1)^{l_\up-l_\down}e^{-ikQ\ell_B^2}}{2}f_Q\Delta.
\end{align}
Therefore, the eigenvalues of the quasiparticle Hamiltonian in Eq.~(\ref{eq:hamiltonian}) are provided by $\pm|\cos(kQ\ell_B^2+\varphi)f_Q\Delta|$ with $\varphi=(l_\up-l_\down)\pi/2$ for each $k$, so that the ground-state energy is
\begin{align}
\frac{E_\LO}{L^2} &= \frac{|\Delta|^2}{2g}
- \frac1{L^2}\sum_k|\cos(kQ\ell_B^2+\varphi)f_Q\Delta| \\
&\!\!\!\!\underset{L\to\infty}\to \frac{|\Delta|^2}{2g}
- \frac{m\omega_B}{\pi^2}|f_Q\Delta|
\end{align}
with the thermodynamic limit taken at fixed $Q\neq0$.
Its minimization with respect to $\Delta$ and $Q$ leads to
\begin{align}\label{eq:energy_LO}
\min\frac{E_\LO}{L^2} = -\frac{g}{2}\left(\frac{m\omega_B}{\pi^2}|f_{\bar Q}|\right)^2,
\end{align}
where $|\Delta|=g\frac{m\omega_B}{\pi^2}|f_{\bar Q}|$ with the same $\bar Q$ maximizing $|f_Q|$.
By comparing Eqs.~(\ref{eq:energy_FF}) and (\ref{eq:energy_LO}), we find
\begin{align}
\min E_\FF = \frac{\pi^2}{8}\min E_\LO < \min E_\LO < 0,
\end{align}
and the FF state thus proves to be energetically favored over the LO state.

\subsection{Zero-temperature phase diagram}
Now that the FF state is energetically favored in the weak-coupling limit, we set $\Delta(\x)=e^{iQy}\Delta$ and elucidate the zero-temperature phase diagram at weak coupling.
To this end, we allow each chemical potential to lie slightly off the Landau level, i.e., $\mu_\sigma=\eps_{l_\sigma}+\delta\mu_\sigma$ with $|\delta\mu_\sigma|\ll\omega_B$, so that the mixing with the other Landau levels is still negligible.
By denoting $\delta\mu_\up\equiv\delta\mu+\delta h$ and $\delta\mu_\down\equiv\delta\mu-\delta h$, the mean-field Hamiltonian in Eq.~(\ref{eq:hamiltonian}) with the use of Eq.~(\ref{eq:FF}) is modified into
\begin{align}
& H_\MF = L^2\frac{|\Delta|^2}{g} - \sum_k(\delta\mu - \delta h) \notag\\
& + \sum_k\tilde\Phi^\+(k)
\begin{pmatrix}
-\delta\mu - \delta h & -e^{ikQ\ell_B^2}f_Q\Delta \\
-e^{-ikQ\ell_B^2}f_Q^*\Delta^* & \delta\mu - \delta h
\end{pmatrix}
\tilde\Phi(k).
\end{align}
Because the eigenvalues of the quasiparticle Hamiltonian are provided by $\pm\sqrt{\delta\mu^2+|f_Q\Delta|^2}-\delta h$ for every $k$, the ground-state energy is
\begin{align}\label{eq:energy_weak-coupling}
\frac{E}{L^2} &= \frac{|\Delta|^2}{g}
- \frac{m\omega_B}{2\pi}\textstyle\bigl(\sqrt{\delta\mu^2+|f_Q\Delta|^2} + \delta\mu\bigr) \notag\\
&\quad - \frac{m\omega_B}{2\pi}\textstyle\bigl(|\delta h| - \sqrt{\delta\mu^2+|f_Q\Delta|^2}\bigr)_>,
\end{align}
where $(x)_>\equiv x\,\theta(x)$ for brevity.

The resulting phases are determined by finding $\Delta$ and $Q$ minimizing the ground-state energy.
It is minimized with respect to $Q$ always at $Q=\bar Q$ maximizing $|f_Q|$.
A phase with $\Delta\neq0$ then corresponds to the unpolarized superfluid (SF) or FF state depending on $l_\up=l_\down$ ($\bar Q=0$) or $l_\up\neq l_\down$ ($\bar Q\neq0$).
On the other hand, a phase with $\Delta=0$ corresponds to the normal state where the system is composed of two quantum Hall insulators (QHIs) with filling factors of $(\nu_\up,\nu_\down)=[l_\up+\theta(\delta\mu_\up),l_\down+\theta(\delta\mu_\down)]$ for two spin components of fermions.
In particular, when $\nu_\up=\nu_\down>0$, the system without population imbalance is the quantum spin Hall insulator with time-reversal invariance~\cite{Kane:2005a,Kane:2005b,Bernevig:2006,Anzai:2017}.

\begin{figure}[t]
\includegraphics[width=0.96\columnwidth]{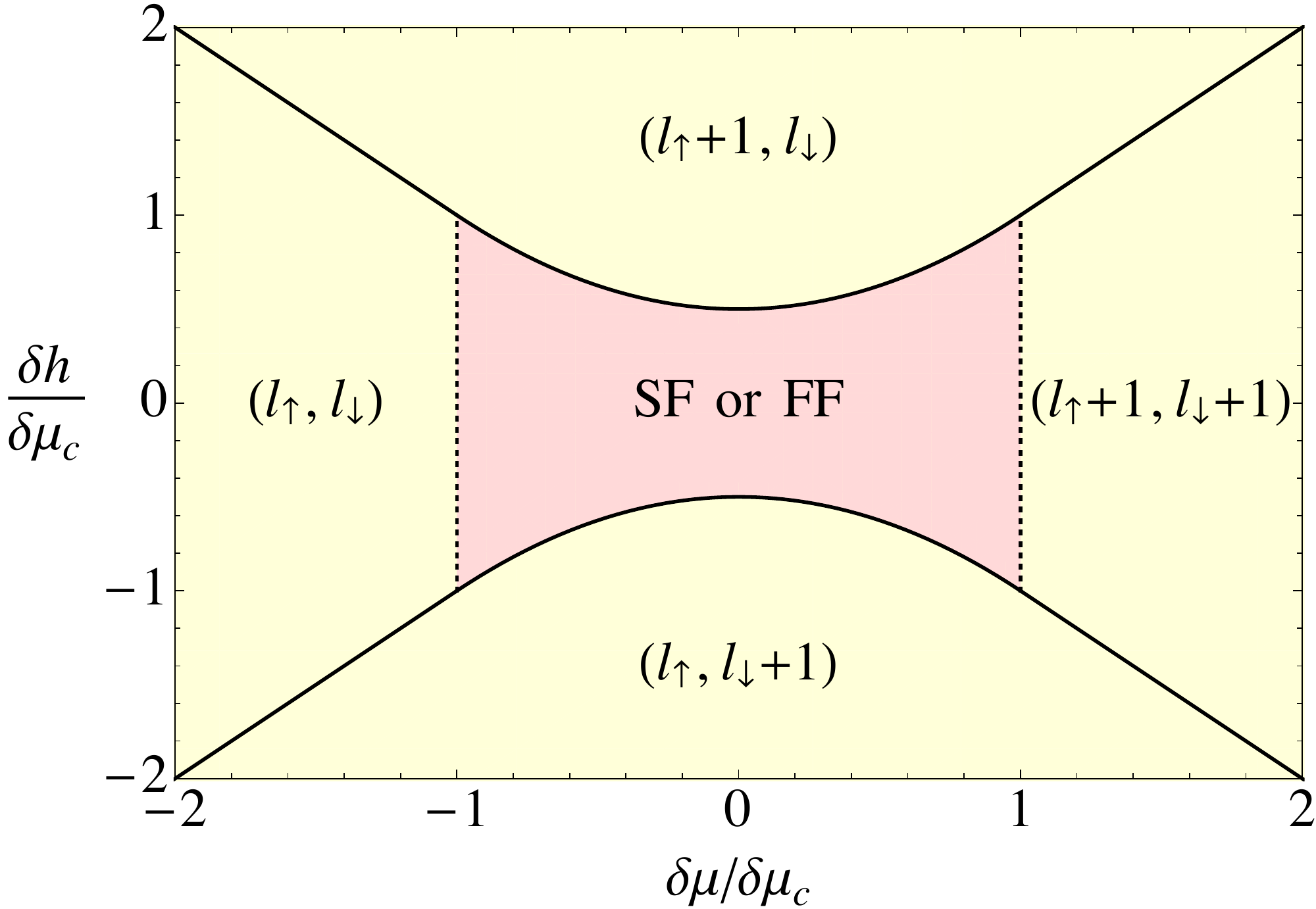}
\caption{\label{fig:weak-coupling_local}
Zero-temperature phase diagram at weak coupling in the plane of $\delta\mu$ and $\delta h$ for $\mu_\sigma\simeq\eps_{l_\sigma}$ in units of $\delta\mu_c$ defined in Eq.~(\ref{eq:dmu_cri}).
The unpolarized SF or FF phase is located at the center depending on $l_\up=l_\down$ or $l_\up\neq l_\down$, and the other four phases are QHIs whose filling factors $(\nu_\up,\nu_\down)$ are indicated.
The quantum phase transitions denoted by the dotted lines at $|\delta\mu|=\delta\mu_c$ are of the second order, whereas those denoted by the solid lines at $|\delta h|=(\delta\mu^2+\delta\mu_c^2)/2\delta\mu_c<\delta\mu_c$ and at $|\delta h|=|\delta\mu|>\delta\mu_c$ are of the first order.}
\end{figure}

It is an elementary analysis to minimize the ground-state energy in Eq.~(\ref{eq:energy_weak-coupling}) with respect to $\Delta$, and the obtained phase diagram is presented in Fig.~\ref{fig:weak-coupling_local} in the plane of $\delta\mu$ and $\delta h$.
When $|\delta\mu|>|\delta h|$, we find $|\Delta|=\sqrt{\delta\mu_c^2-\delta\mu^2}/|f_{\bar Q}|$ for $|\delta\mu|<\delta\mu_c$ and $\Delta=0$ for $|\delta\mu|>\delta\mu_c$, which are separated by a second-order quantum phase transition at
\begin{align}\label{eq:dmu_cri}
|\delta\mu| = \delta\mu_c \equiv g\frac{m\omega_B}{4\pi}|f_{\bar Q}|^2.
\end{align}
On the other hand, when $|\delta h|>|\delta\mu|$, we find $\Delta=0$ for $|\delta h|>\delta\mu_c$ and for $\delta h_c<|\delta h|<\delta\mu_c$, whereas $|\Delta|=\sqrt{\delta\mu_c^2-\delta\mu^2}/|f_{\bar Q}|$ for $|\delta h|<\delta h_c,\delta\mu_c$.
They are separated by a first-order quantum phase transition at
\begin{align}
|\delta h| = \delta h_c \equiv \frac{\delta\mu^2 + \delta\mu_c^2}{2\delta\mu_c} < \delta\mu_c,
\end{align}
and it continues into another first-order quantum phase transition at $|\delta h|=|\delta\mu|>\delta\mu_c$ separating the two QHI phases with different filling factors.
The quasiparticle energy in the SF or FF phase is thus found to be $\sqrt{\delta\mu^2+|f_Q\Delta|^2}=\delta\mu_c$ independent of $\delta\mu$, and the ratio $\delta h_c/\delta\mu_c$ varies from $1/2$ to $1$ for $0\leq|\delta\mu|\leq\delta\mu_c$.

\begin{figure}[t]
\includegraphics[width=0.96\columnwidth]{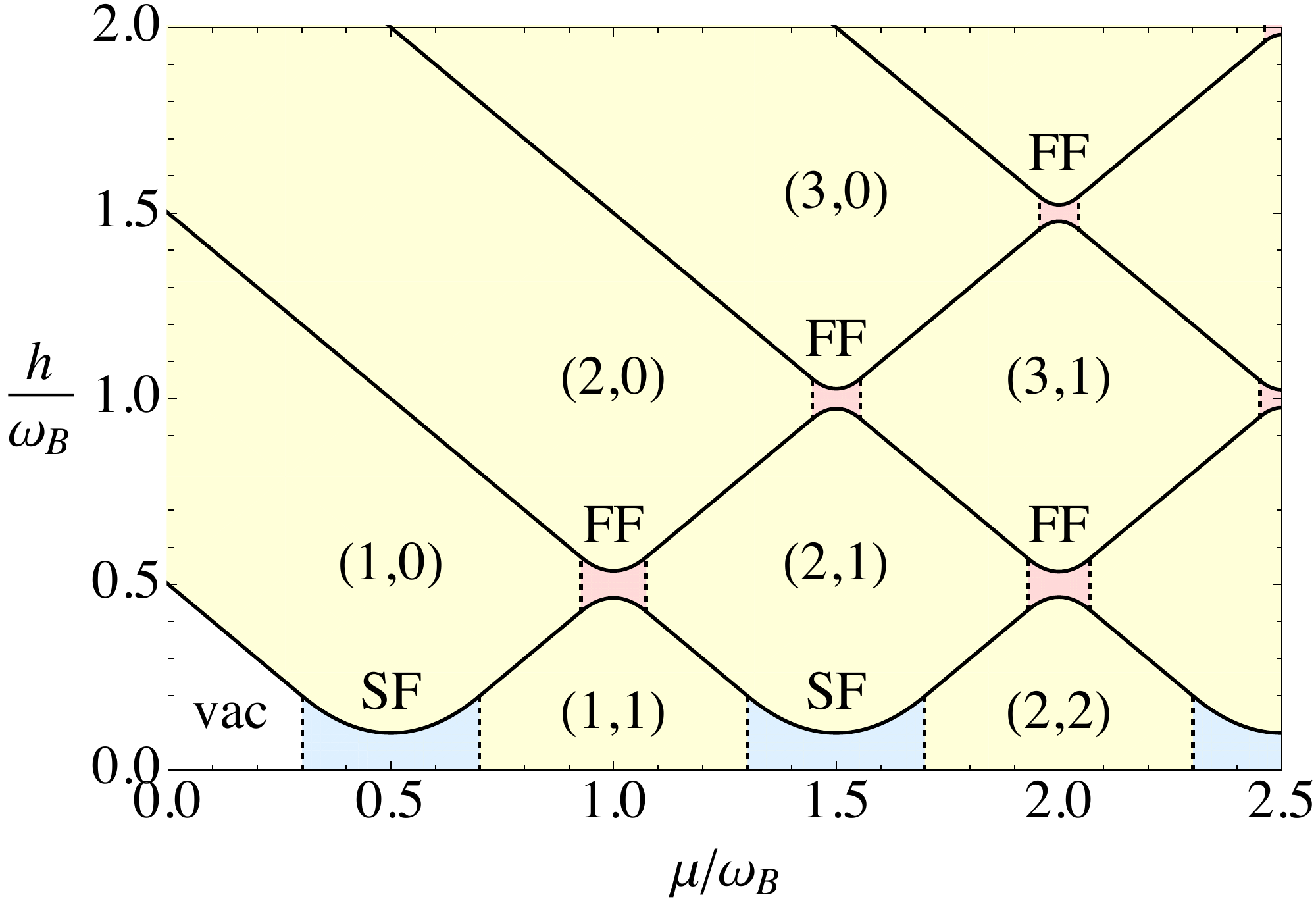}
\caption{\label{fig:weak-coupling_global}
Zero-temperature phase diagram in the plane of the average chemical potential $\mu=(\mu_\up+\mu_\down)/2$ and the Zeeman field $h=(\mu_\up-\mu_\down)/2$ in units of the cyclotron frequency $\omega_B$ deduced from Fig.~\ref{fig:weak-coupling_local} for $mg=2.5$.
 See the caption of Fig.~\ref{fig:weak-coupling_local} for the other details.}
\end{figure}

The above phase diagram is valid in the vicinity of $\mu_\sigma=\eps_{l_\sigma}$ for every $l_\sigma=0,1,2,\dots$, which all together leads to the global phase diagram in the plane of the average chemical potential $\mu\equiv(\mu_\up+\mu_\down)/2$ and the Zeeman field $h\equiv(\mu_\up-\mu_\down)/2$ as presented in Fig.~\ref{fig:weak-coupling_global}.
It consists of the QHI, SF, and FF phases as well as the vacuum ($\nu_\up=\nu_\down=0$) and is symmetric under $h\to-h$ with $\nu_\up\leftrightarrow\nu_\down$.
The SF or FF and QHI phases are separated by the second-order (first-order) quantum phase transition along the direction of $\mu$ ($h$), whereas the two QHI phases with different filling factors $(\nu_\up,\nu_\down)$ are separated by the first-order quantum phase transition.
We note that a large value of $mg=2.5$ is artificially chosen for Fig.~\ref{fig:weak-coupling_global} in order to make the FF phases visible.

\section{Phase diagram beyond weak coupling}\label{sec:beyond}
Because the FF phases occupy only the tiny portions of the phase diagram in the weak-coupling limit, it is important to elucidate how they extend beyond it.
To this end, we go back to the mean-field Hamiltonian in Eq.~(\ref{eq:mean-field}) and expand the fermion field operator over all the eigenfunctions in Eq.~(\ref{eq:eigenfunction}) so as to allow for the mixing of the Landau levels.
By assuming the FF ansatz, $\Delta(\x)=e^{iQy}\Delta$, making the gauge transformation, $\phi_\sigma(\x)\to e^{iQy/2}\phi_\sigma(\x)$, and then substituting $\phi_\up(\x)=\sum_{k,l}\chi_{kl}(\x)\tilde\phi_\up(k,l)$ and $\phi_\down(\x)=\sum_{k,l}\chi_{kl}^*(\x)\tilde\phi_\down(k,l)$ into Eq.~(\ref{eq:mean-field}), the mean-field Hamiltonian reads
\begin{align}
& H_\MF = L^2\frac{|\Delta|^2}{g}
+ \sum_{k,l}(\eps_l - \mu + h) \notag\\
& + \sum_{k,l,l'}\tilde\Phi^\+(k,l)\left[
\begin{pmatrix}
\eps_l + \frac{Q^2}{8m} - \mu - h & -\Delta \\
-\Delta^* & -\eps_l - \frac{Q^2}{8m} + \mu - h
\end{pmatrix}
\delta_{ll'}\right. \notag\\
& + \left.\frac{iQ}{2m\ell_B}
\begin{pmatrix}
1 & 0 \\
0 & 1
\end{pmatrix}
\left(\sqrt{\frac{l}{2}}\,\delta_{l-1,l'}
- \sqrt{\frac{l+1}{2}}\,\delta_{l+1,l'}\right)\right]\tilde\Phi(k,l').
\end{align}
Here, the last term is the quasiparticle Hamiltonian in the Nambu-Gor'kov basis with $\tilde\Phi(k,l)\equiv[\tilde\phi_\up(k,l),\tilde\phi_\down^\+(k,l)]^T$, and we employed $\ell_BF'_l(y)=\sqrt{l/2}\,F_{l-1}(y)-\sqrt{(l+1)/2}\,F_{l+1}(y)$.
Because the eigenvalues of the quasiparticle Hamiltonian are provided in the forms of $\pm E_l-h$ ($E_l>0$; $l=0,1,2,\dots$) for every $k$, the ground-state energy is
\begin{align}\label{eq:energy_beyond}
\frac{E}{L^2} &= \frac{|\Delta|^2}{g}
- \frac{m\omega_B}{2\pi}\sum_l(E_l - \eps_l + \mu)\,\theta(\Lambda - \eps_l) \notag\\
&\quad - \frac{m\omega_B}{2\pi}\sum_l(|h| - E_l)_>.
\end{align}

The quasiparticle energy $E_l$ depends on $\mu$, $\Delta$, and $Q$ and is confirmed to have the asymptotic form of $\lim_{l\to\infty}E_l=\sqrt{(\eps_l-\mu)^2+|\Delta|^2}$.
Therefore, the second term in the right-hand side of Eq.~(\ref{eq:energy_beyond}) is logarithmically divergent and is regularized by introducing an energy cutoff $\Lambda$.
This logarithmic divergence should be canceled by the same form of divergence hidden in the coupling constant~\cite{Marini:1998},
\begin{align}\label{eq:coupling}
\frac1g = \frac{m}{2\pi}\int_0^\Lambda\!\frac{d\epsilon}{2\eps+\eps_b}.
\end{align}
Here, $\eps_b=2\Lambda\exp(-4\pi/mg)>0$ has the physical meaning of the binding energy of a two-body bound state in the vacuum without magnetic fields, which always exists for any $g>0$ in 2D and can thus be used to parametrize the attraction~\cite{Randeria:1989,Randeria:1990}.
By separating out the divergent piece from the second term, combining it with the first term, and then taking the limit of $\Lambda\to\infty$, the ground-state energy in Eq.~(\ref{eq:energy_beyond}) is made manifestly cutoff independent as
\begin{align}
& \frac{E}{L^2} = \frac{m|\Delta|^2}{4\pi}\left[\ln\!\left(\frac{2\omega_B}{\eps_b}\right)
+ \psi\!\left(\frac12 - \frac\mu{\omega_B}\right)\right] \notag\\
& - \frac{m\omega_B}{2\pi}\sum_l\left[E_l - \eps_l + \mu - \frac{|\Delta|^2}{2(\eps_l - \mu)} + (|h| - E_l)_>\right],
\end{align}
where $\psi(z)$ is the digamma function~\cite{Anzai:2017}.

\begin{figure*}[p]
~\!\includegraphics[width=0.98\columnwidth]{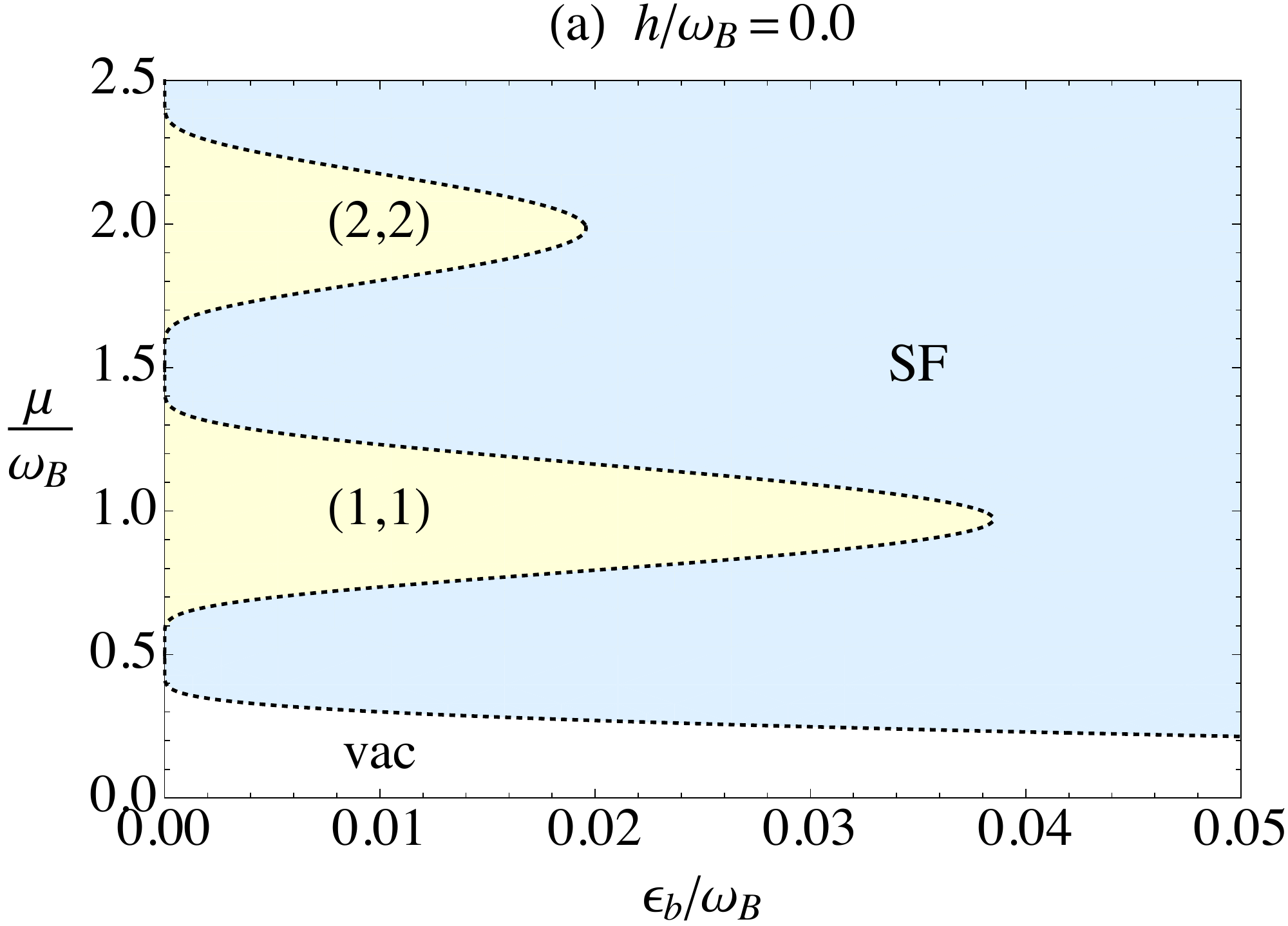}\hfill
\includegraphics[width=0.965\columnwidth]{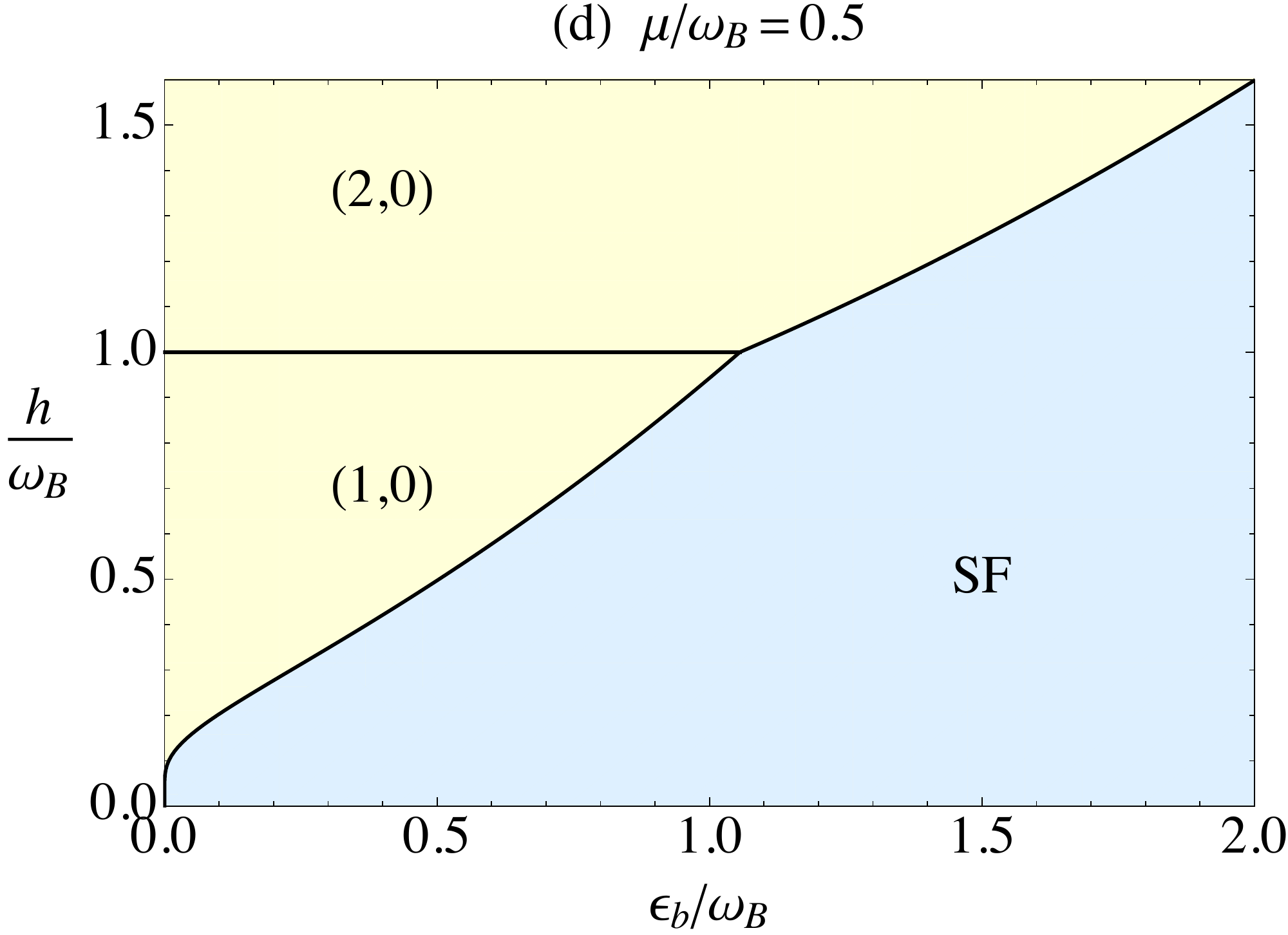}\\\bigskip\bigskip
\includegraphics[width=0.965\columnwidth]{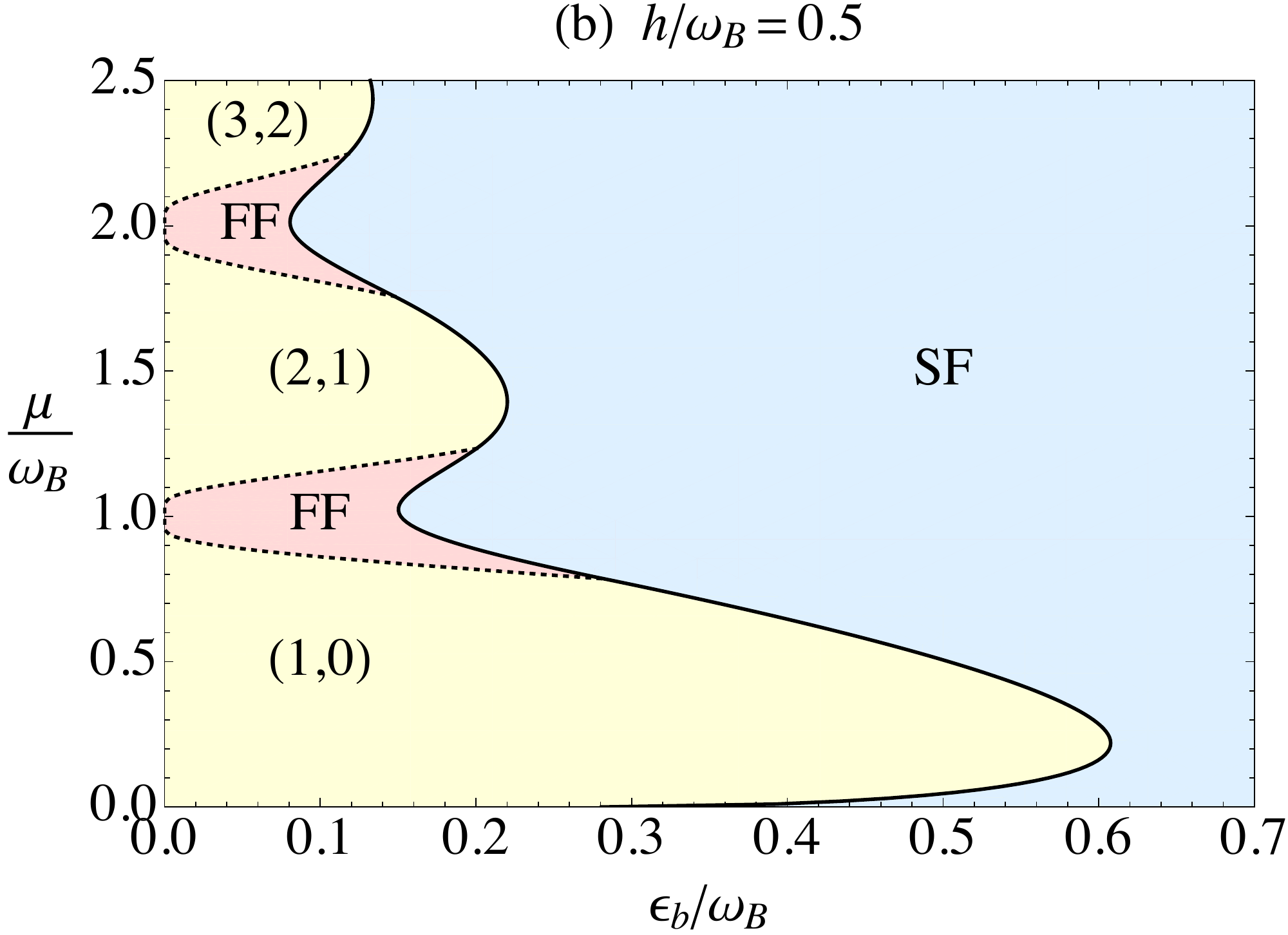}\hfill
\includegraphics[width=0.965\columnwidth]{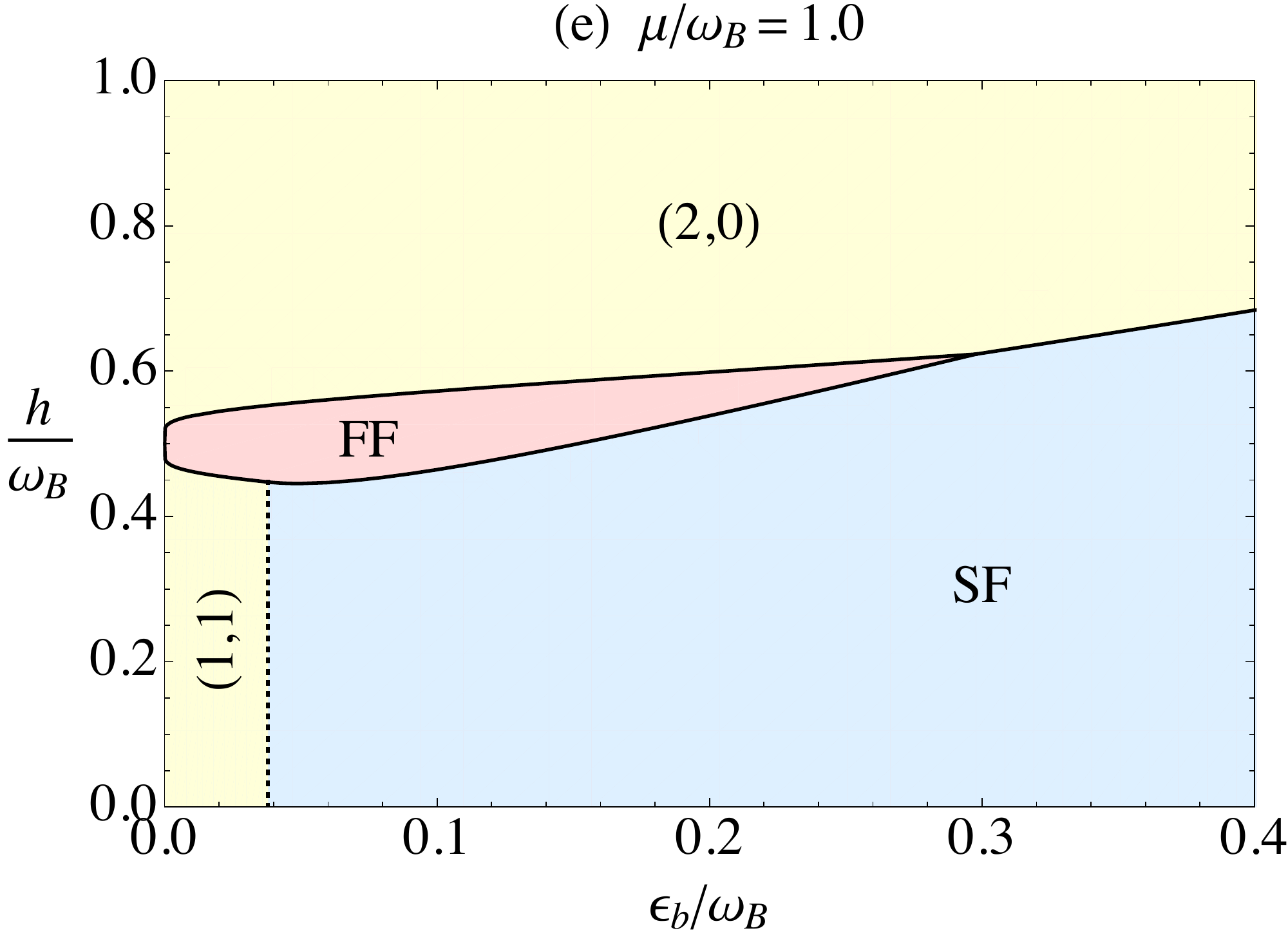}\\\bigskip\bigskip
\includegraphics[width=0.94\columnwidth]{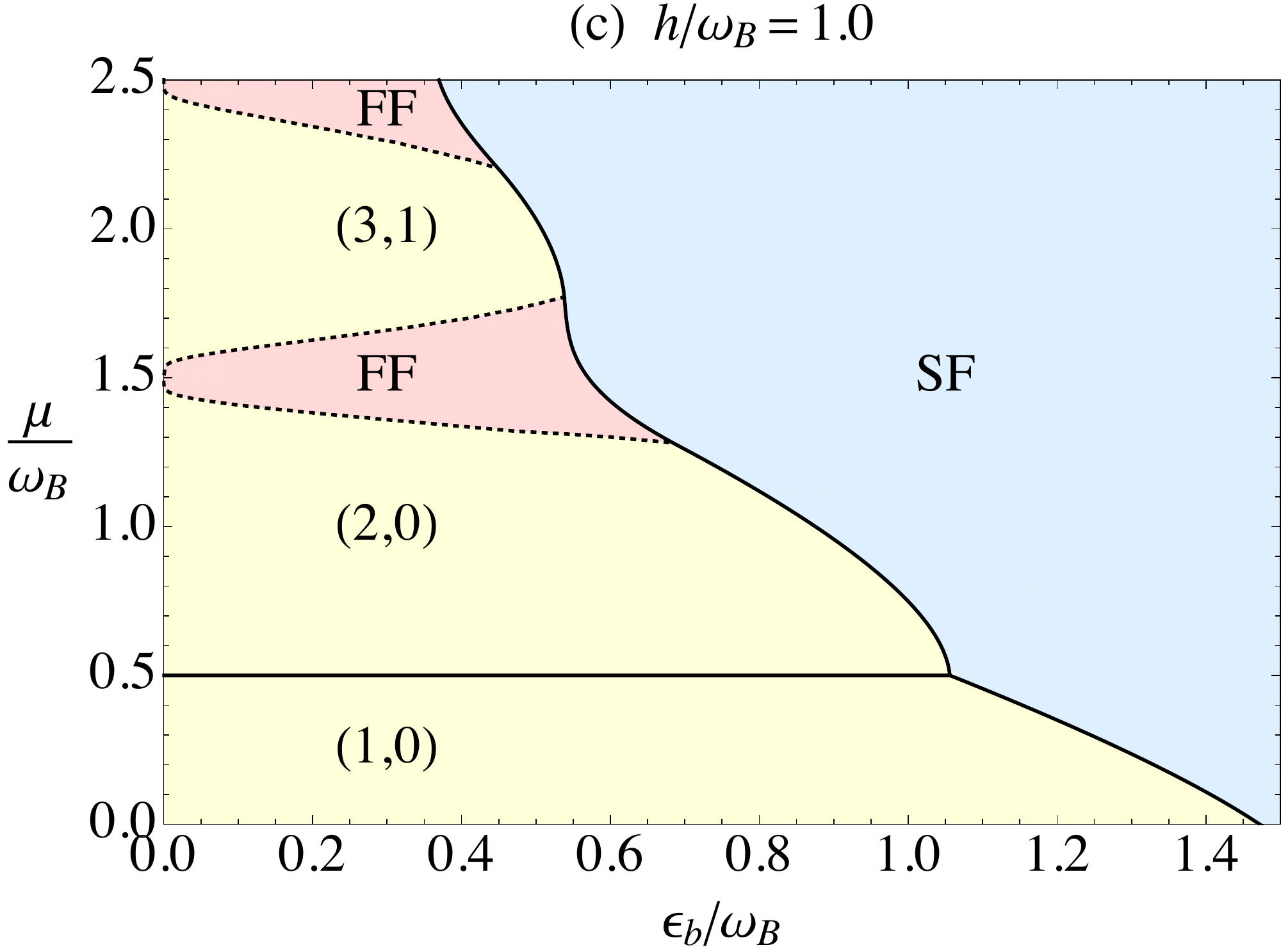}\!~~\hfill
\includegraphics[width=0.965\columnwidth]{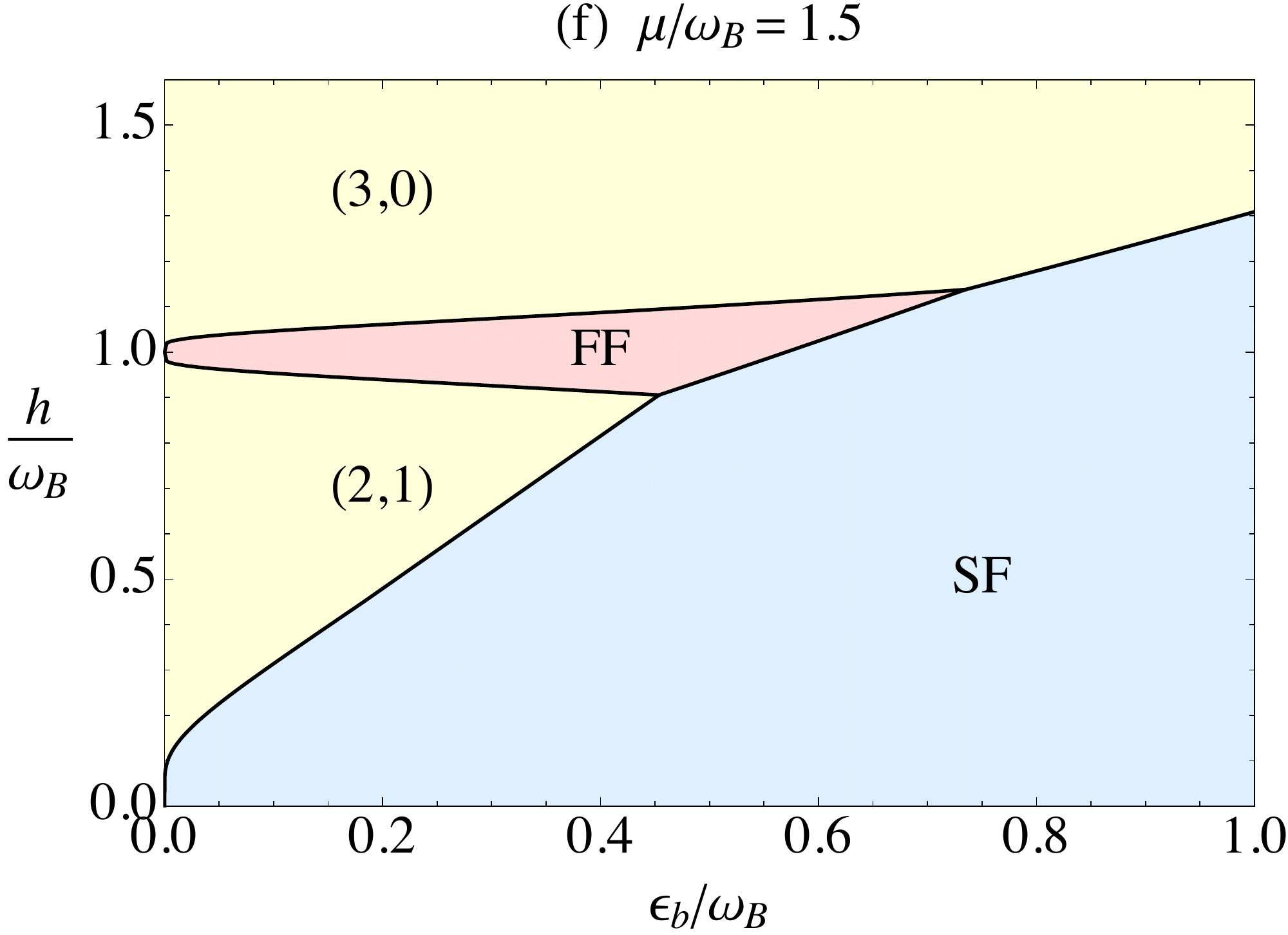}\\\bigskip
\caption{\label{fig:beyond}
Zero-temperature phase diagrams in the planes of the two-body binding energy $\eps_b$ and the average chemical potential $\mu$ at (a) $h/\omega_B=0.0$~\cite{Anzai:2017}, (b) $0.5$, and (c) $1.0$ and in the planes of $\eps_b$ and the Zeeman field $h$ at (d) $\mu/\omega_B=0.5$, (e) $1.0$, and (f) $1.5$ in units of the cyclotron frequency $\omega_B$.
There appear the unpolarized SF and FF phases as well as QHIs with their filling factors $(\nu_\up,\nu_\down)$ indicated.
They are separated by the various second-order (dotted curves) and first-order (solid curves) quantum phase transitions.}
\end{figure*}

We minimize the resulting ground-state energy numerically to find $\Delta$ and $Q$ by employing up to $l_{\max}=500$ quasiparticle energies, which are confirmed to be enough for convergence of the summation over $l$.
The obtained phase diagrams are presented in Fig.~\ref{fig:beyond} in the planes of $\eps_b$ and $\mu$ at (a) $h/\omega_B=0.0$~\cite{Anzai:2017}, (b) $0.5$, and (c) $1.0$ and in the planes of $\eps_b$ and $h$ at (d) $\mu/\omega_B=0.5$, (e) $1.0$, and (f) $1.5$.
They consist of the QHI ($\Delta=0$), SF ($\Delta\neq0$, $Q=0$), and FF ($\Delta\neq0$, $Q\neq0$) phases and are consistent with the phase diagram at weak coupling elucidated in the previous section.
In particular, we find that each FF phase appearing at $\mu=(\eps_{l_\up}+\eps_{l_\down})/2$ and $h=(\eps_{l_\up}-\eps_{l_\down})/2$ with $l_\up\neq l_\down$ in the weak-coupling limit extends well by increasing the attraction and is eventually replaced by the SF phase with the first-order quantum phase transition.
The QHI phase is also found to be replaced by the SF phase with the second-order (first-order) quantum phase transition for $\nu_\up=\nu_\down$ ($\nu_\up\neq\nu_\down$) by increasing the attraction.
Because the FF phases prove to occupy the reasonable portions of the phase diagram, they may, in principle, be realized by ultracold atom experiments.

\section{Conclusion and outlook}
We studied a 2D Fermi gas with an attractive interaction in antiparallel magnetic fields with population imbalance.
By employing the mean-field approximation, we showed that the FF state is energetically favored over the LO state in the weak-coupling limit.
We then elucidated the zero-temperature phase diagram in the space of attraction, average chemical potential ($\mu$), and Zeeman field ($h$) analytically at weak coupling (see Fig.~\ref{fig:weak-coupling_global}) as well as numerically beyond it (see Fig.~\ref{fig:beyond}).
It was found to show the rich structures consisting of QHI, unpolarized SF, and FF phases, where
\begin{enumerate}\renewcommand\labelenumi{(\roman{enumi})}
\setlength\itemindent{0pt}\setlength\itemsep{0pt}
\item the SF or FF and QHI phases are separated by the second-order (first-order) quantum phase transition along the direction of $\mu$ ($h$),
\item the two QHI phases with different filling factors $(\nu_\up,\nu_\down)$ are separated by the first-order quantum phase transition,
\item the FF phase is replaced by the SF phase with the first-order quantum phase transition by increasing the attraction,
\item the QHI phase is replaced by the SF phase with the second-order (first-order) quantum phase transition for $\nu_\up=\nu_\down$ ($\nu_\up\neq\nu_\down$) by increasing the attraction.
\end{enumerate}

In particular, the FF phases proved to occupy the reasonable portions of the phase diagram, so that they may, in principle, be realized by ultracold atom experiments.
To this end, it is worthwhile to note that the average chemical potential for trapped systems is replaced by $\mu(\x)=\mu_0-V(\x)$ within the local-density approximation, whereas $h$ remains constant~\cite{Bloch:2008}.
Therefore, by realizing a sufficiently large $\mu_0$ with $|h|\simeq\N_+\omega_B/2$, a series of FF phases separated by QHIs may appear along the path from the trap center $\mu(\0)=\mu_0$ towards the edge $\mu(|\x|\to\infty)\to-\infty$.

As for future works, we plan to extend our study to finite temperature where fluctuations beyond the mean-field approximation need to be taken into consideration.
It is also interesting to extend our study to 3D with population imbalance where different types of FF states are possible, such as the Cooper pairing with nonzero momentum parallel to magnetic fields, perpendicular to magnetic fields, and both of them.

\acknowledgments
The authors thank Shunji Tsuchiya and Ryosuke Yoshii for valuable discussions.
This work was supported by JSPS KAKENHI Grants No.~JP15K17727 and No.~JP15H05855.

\end{document}